\documentclass[10pt, a4paper]{article} 

\usepackage{graphicx, caption}%
\usepackage{amsmath,amssymb,amsfonts}%
\usepackage{float}

\begin{document}

\title{Diffusion-based kernel density estimation improves the assessment of carbon isotope modelling}
\author{Maria-Theresia Pelz \footnote{Kiel University, Department of Computer Science}
\and Christopher Somes \footnote{GEOMAR Helmholtz Centre for Ocean Research Kiel, Biogeochemical Modelling}}
\date{2023 \\ August}

\maketitle

\abstract{Comparing differently sized data sets is one main task in model assessment and calibration. This is due to field data being generally sparse compared to simulated model results. We tackled this task by the application of a new diffusion-based kernel density estimator (diffKDE) that approximates probability density functions of a data set nearly independent of the amount of available data. We compared the resulting density estimates of measured and simulated marine particulate organic  carbon-13 isotopes qualitatively and quantitatively by the Wasserstein distance. For reference we also show the corresponding comparison based on equally sized data set with reduced simulation and field data. The comparison based on all available data reveals a better fit of the simulation to the field data and shows misleading model properties in the masked analysis. A comparison between the diffKDE and a traditional Gaussian KDE shows a better resolution of data features under the diffKDE. We are able to show a promising advantage in the application of KDEs in calibration of models, especially in the application of the diffKDE.}


\section{Introduction}\label{sec1}
Ocean data are highly diverse and thus require good performance of evaluation tools on different data features. They can describe individual biological, chemical or geological tracers, resolve physical properties of the ocean, be linked to specific location or time or even to each other. Sources of marine data furthermore diversify marine data ~\cite{MartnMguez2019, Schmidt2020}. They can be collected by field measurements from a research vessel, time series stations, autonomous devices or fixed traps in the water column. Furthermore, they can be obtained as results from simulations of marine processes as for example included in Earth system models. Being influenced by various processes, marine data is often multimodal ~\cite{Burian2021, Hildebrand2022}, sometimes boundary close ~\cite{Lampe2021} and generally noised ~\cite{Chen2020, Wang2018, Gilkeson2014}. Hence, a tool for the evaluation of marine data must account for this specific characteristics and resolve the true data structure under the noise.

Comparing marine data is a fundamental task in ocean research. It can assess changes in measurement data, to evaluate projections and test cases from simulations and to assess the quality of a model ~\cite{Sfrian2012}. In many cases it is important to be able to compare differently sized data sets. In model assessment and calibration, sparsity of field data induces the need to compare differently sized data sets. Field data is generally only available in measured times and locations, whereas simulation data exists in every grid cell and for every time step. To compare such differently sized data, a mask can be applied to reduce the data to a comparable amount. In model assessment this mask is generally chosen to mark the grid cells, where both data kinds are available and only incorporate data from these grid cells ~\cite{Somes2021}.

Approximated probability density functions (PDFs) allow to investigate data nearly independent of their size and by this build a basis to compare differently sized data ~\cite{Sheather2004}. PDFs give an intuitive visual insight into the distribution of data and provide a continuous function for following analyses. There are two main approaches towards the estimation of PDFs: parametric and non-parametric ~\cite{Tsybakov2009}. The parametric approach assumes the knowledge of an underlying specific density and aims at estimating its parameters. This can be an efficient way for approximation, but requires the assumption of an underlying specif density to be true and hence is generally insufficient for such diverse data as from the marine environment. The non-parametric approximation does not require any knowledge about the input data and attempts to estimate the density by weighing all input data equally. This offers the opportunity to explore data with multiple modes of unknown count and location as in many marine data.

The most prominent non-parametric PDF estimator is a kernel density estimator (KDE) ~\cite{Sheather2004}. There exist a variety of choices for the KDEs. A common choice is a Gaussian KDE build on the distribution function of the Gaussian distribution. Unfortunately, this tends to oversmooth multimodal structure and is also inconsistent at the boundaries of restricted domains ~\cite{Botev2010}. An improved approach on this specific tasks in a diffusion-based KDE ~\cite{Botev2010} build on the solution of the diffusion heat equation. 

We used a new implementation of a diffusion-based KDE (diffKDE) to show a new comparison possibility for simulated and measured marine particulate organic carbon-13 isotopes, expressed in the delta notation which is based on its ratio relative to carbon-12  ($\delta^{13}$C$_{\text{POC}}$). The diffKDE includes optimal smoothing properties for geoscientific data and well resolves typical structures of marine data ~\cite{Pelz2023}. Simulation results are obtained by  ~\cite{Somes2021} and corresponding field data by ~\cite{pelz2021data}. We created two test scenarios comparing (1) a traditional masked data approach with the comparison of equally size data and (2) a full data approach with the comparison of all available data. 

The paper is structured as followed: In the second section, we describe the method of kernel density estimation and introduce the two estimators applied in this study. Furthermore, we describe the used example data of marine carbon-13 isotopes. The third section shows the results of the two estimators on the test data in the two scenarios and four test cases each. The paper ends with a short discussion of the observed results and a concluding section.

\section{Methods}\label{sec2}

We applied the non-parametric approach of a kernel density estimator (KDE) to approximate probability density functions (PDFs) of carbon isotope data. This allows for the comparison of differently sized data by switching from a direct data comparison to a comparison of their densities. The resulting estimates from simulated and measured field data can be compared qualitatively by eye and quantitatively by a divergence function ~\cite{Thorarinsdottir2013}. In the following we shortly describe the applied KDEs and the incorporated data.

\subsection{Kernel density estimation by diffusion and Gaussian kernels}\label{subsec2}

A very common choice for a KDE is a weighed sum of Gaussian kernels ~\cite{Sheather2004}. This sum definition of a KDE can generally be given as
\begin{equation}
\label{eq:kde}
\hat{f}: \mathbb{R} \times \mathbb{R}_{>0} \rightarrow \mathbb{R}_{\geq 0},
\quad\left(x ; t \right) \mapsto\frac{1}{n \, \sqrt{t}} \sum _{j = 1} ^n  K \left( \frac{x - X_j}{\sqrt{t}} \right).
\end{equation}
~\cite{parzen}. For the Gaussian KDE the kernel function $K:\mathbb{R} \rightarrow \mathbb{R}_{>0}$ is set to be the density of the Gaussian distribution as
\begin{equation}\label{eq:gauss}
\Phi: \mathbb{R} \rightarrow \mathbb{R}_{\geq 0},  w \mapsto \frac{1}{\sqrt{2 \pi}} e^{-\frac{1}{2} w^2}.
\end{equation}
The here applied Gaussian KDE is part of the stats package from the SciPy Python library ~\cite{GommersScipy}.

\begin{figure}[h]%
\centering
\includegraphics[width=0.9\textwidth]{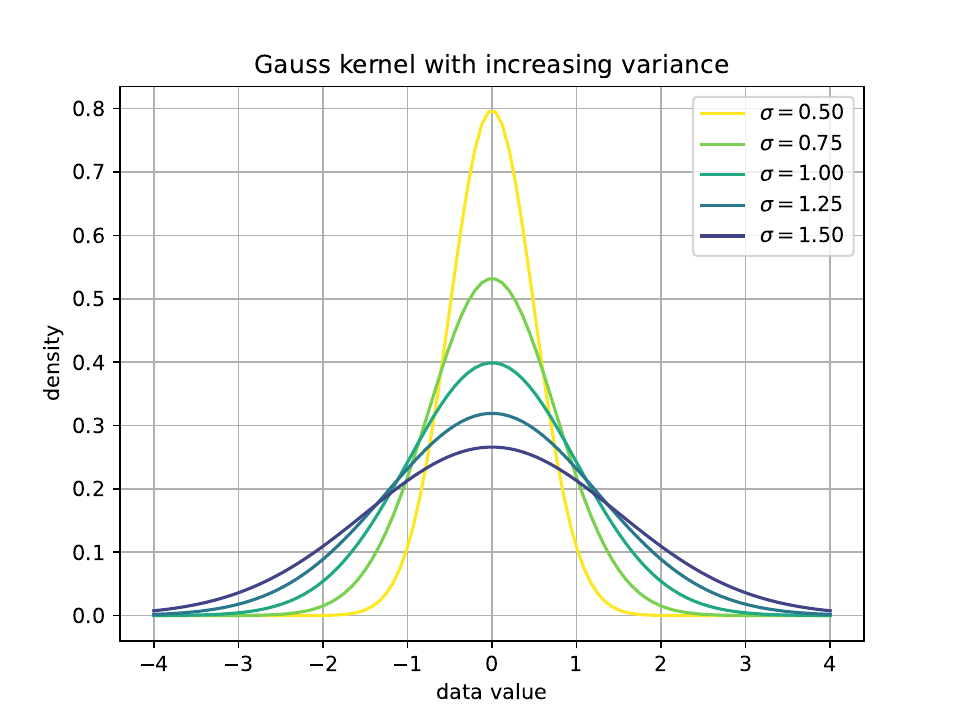}
\caption{A Gauss kernel centered around $0$ with increasing variance from $\sigma = 0.5$ to $\sigma=1.5$.}\label{fig2}
\end{figure}

Despite widely applied, the Gaussian KDE has several disadvantages in the application on marine data. These include inconsistency at the boundaries of restricted data domains and oversmoothing of multimodal distributions ~\cite{Botev2010}. To better account for these shortcomings, we expanded the Gaussina KDE approach to the diffusion-based KDE (diffKDE). The here applied diffKDE was proposed by ~\cite{Chaudhuri2000}, expanded by ~\cite{Botev2010} and implemented by ~\cite{Pelz2023} in ~\cite{Pelz2023KDE}. The motivation originates from the Gaussian distribution being closely related to the diffusion process. An increase in the variance parameter $t\in\mathbb{R}_{>0}$ in Eq. ~\ref{eq:kde} with the Gaussian kernel from Eq. ~\ref{eq:gauss} can be interpreted with an increase in time while solving the diffusion equation. This thought experiment is visualized in Fig. ~\ref{fig2}. Mathematically, the correlation between Gaussian KDE and diffusion equation is given by the Gaussian kernel solving the diffusion equation as a fundamental solution ~\cite{Chaudhuri2000}. 

The diffKDE is defined as the solution $u \in C ^{2,1} \left(  \Omega \times \mathbb{R}_{>0} , \mathbb{R}_{\geq 0} \right)$, which solves the diffusion partial differential equation
\begin{align}
\frac{\partial}{\partial t} u\left( x; t \right) &=  \frac{1}{2} \frac{d^2}{dx^2}\left( \frac{u\left( x ; t \right)}{p\left(x\right)} \right), & x \in \Omega, t \in \mathbb{R}_{>0}, \label{eq:eq01} \\
\frac{\partial}{\partial x} \left( \frac{u \left( x;t \right)}{p\left( x \right)} \right) &= 0, & x \in \partial \Omega , t \in \mathbb{R}_{>0}, \label{eq:eq2} \\ 
u\left( x; 0 \right) &= \frac{1}{N} \sum _{j=1}^N \delta \left( x -X_j \right), & x \in \Omega. \label{eq:eq3}
\end{align}
up to a final iteration time $T\in\mathbb{R}_{>0}$. We chose this approach, since the diffKDE promises better results on typical marine data properties. Due to the Neumann boundary conditions it is consistent at the boundaries. The incorporated parameter function $p \in C^2 \left( \Omega, \mathbb{R}_{>0} \right)$ induces adaptive smoothing, which leads to a better resolution of multiple and boundary close modes. The theoretical properties of the diffKDE are in detail discussed in ~\cite{Botev2010}. The here employed implementation, its underlying algorithm for the solution of the partial differential equation in Eq. ~\ref{eq:eq01} and the specific choices for $p$ and $T$ are in detail explained in ~\cite{Pelz2023}. The applied software is available at ~\cite{Pelz2023KDE}.

\begin{figure}[h]%
\centering
\includegraphics[width=0.9\textwidth]{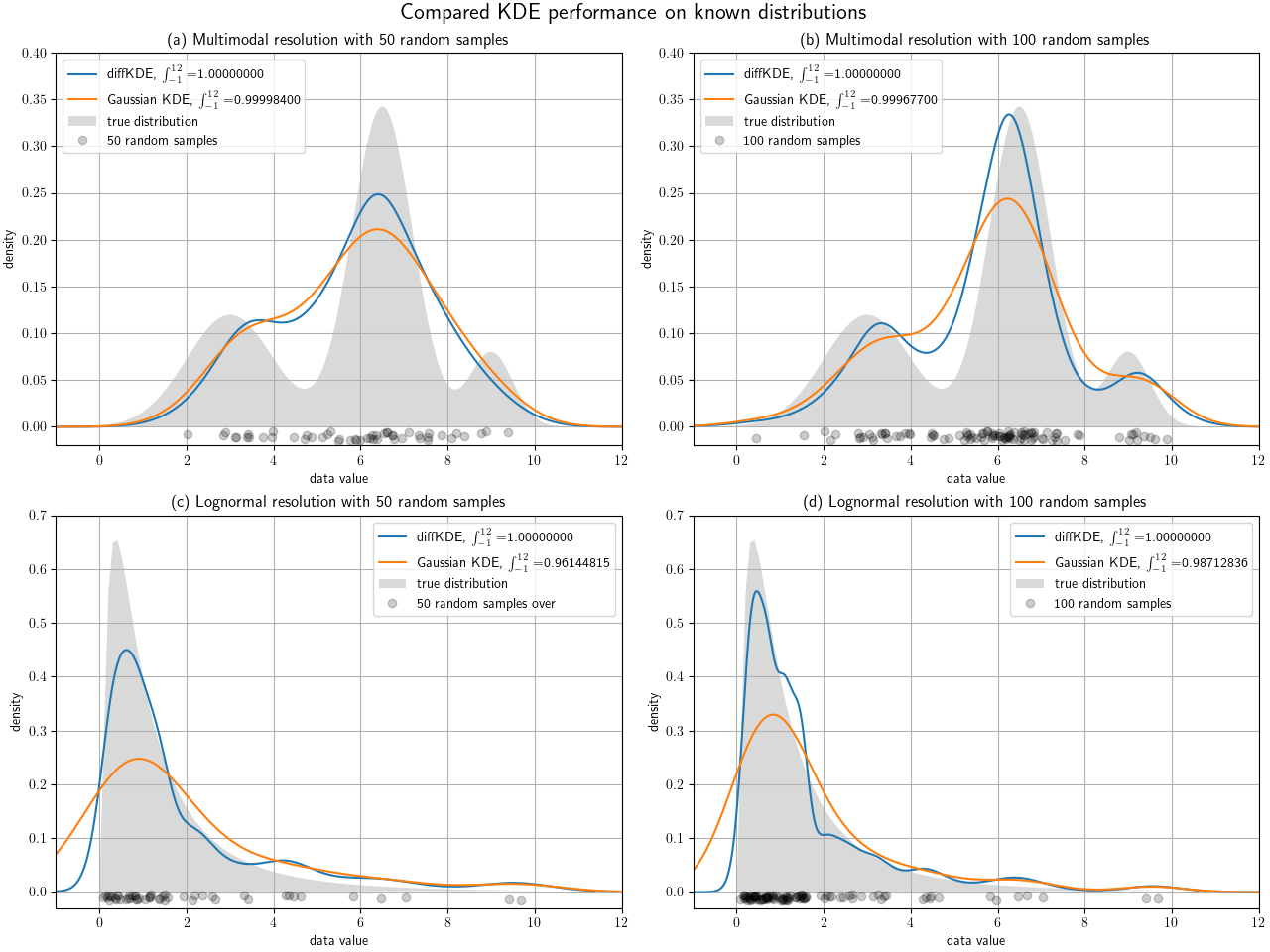}
\caption{Gaussian and diffKDE performance on known data. This figure is an adapted version from ~\cite{Pelz2023}. The known distribution is shown as the grey area in the background. The upper row of panels uses an artificial trimodal distribution. The lower row shows a lognormal distribution. All KDEs are calculated over the domain $\left[-1,12\right]$. The upper and lower left panels show the diffKDE in blue and the Gaussian KDE in orange on a random sample of $50$ data points of the distribution in the background. The upper and lower right show equally the diffKDE and Gaussian KDE on a random sample of $100$ data points of the distribution in the background.}\label{fig2b}
\end{figure}

To provide an insight into the different performance on typical data structures, we used artificial random samples from known distributions and applied both KDEs. The resulting KDEs are shown in comparison to the true distribution in Fig. ~\ref{fig2b} in an adapted graphic from ~\cite{Pelz2023}. In comparison with the Gauss KDE this algorithm shows superior resolution of multimodal and boundary close data. For the multimodal data the Gaussian KDE hardly detects the third mode in both observed cases. Furthermore, the pronounciation of the main mode is fairly undererstimated by the Gaussian KDEin comparison to the diffKDE. The latter detects two modes for the case of 50 random samples and all three modes for 100 random samples. The main mode is well met for the larger random sample. The minimal between the modes are always better resolved by the diffKDE than the Gaussian KDE. In the test cases with the boundary close data the diffKDE detects the height of the mode far better in both test cases than the Gaussian KDE. Furthermore, the steep decline left of the mode is far better met by the diffKDE than the Gaussian KDE, which does not approach zero within the observed domain. This results also in an integral not equal to one for the Gaussian KDE, whereas the diffKDE integrates to one in all observed cases.

\subsection{Carbon-13 isotope data for comparison}

The simulation results are obtained by ~\cite{Somes2021}. This model is built on the UVic Earth System Climate Model version 2.9 ~\cite{Eby2009}. The general circulation has a $1.8^{\circ}\times3.6^{\circ}$ resolution and 19 vertical layers increasing with depth. The biogeochemical model is the Model of Ocean Biogeochemistry and Isotopes (MOBI) version 2.0. This simulates latest findings of carbon cycling ~\cite{Kvale2015} and carbon isotopes ~\cite{Schmittner2016}.

The field data are globally available carbon-13 isotope $\delta^{13}$C$_{\text{POC}}$ by ~\cite{pelz2021data}. The data covers the 1960s to the 2010s and all major ocean basins. A detailed description of the data set version is available at ~\cite{pelz2021}. The data were interpolated onto the grid of the simulation model to make them comparable ~\cite{pelz2021}.

For the scenario (1) of the comparison of equally sized data sets, we applied a mask to both, simulation and field data. This mask marks the grid cells, where both data kinds are available, and only data from these grid cells will be incorporated in the analyses for scenario (1).

Scenario (2) on the other hand will incorporate all available data of the chosen time and domain.

\section{Results}\label{sec3}

We show model assessment approach based on comparison of estimated densities. The simulation and field data are averaged $\delta^{13}$C$_{\text{POC}}$ data over the 1990s. We compare two scenarios: (1) a masked data approach only incorporating data from grid cells, where both data kinds \- simulation and field \- are available and (2) a full data approach incorporating all available data of the chosen time frame and areas. For each of the scenarios, we show four test cases: a comparison of all available data over all depths and ocean basins, a comparison of the data restricted to the euphotic zone, a comparison of the data restricted to the euphotic zone excluding the Southern Ocean and a comparison of the data restricted to the euphotic zone and the Southern Ocean. For all four cases we show comparison of simulation and field data by the diffKDE as well as the traditional Gaussian KDE. We also add to each comparison an error calculated between the graphs for simulation and field data measured by the Wasserstein distance ~\cite{Panaretos2019}. The here discussed test cases are refined experiments already teased in ~\cite{PelzDissertation}.

\begin{figure}[h]%
\centering
\includegraphics[width=0.9\textwidth]{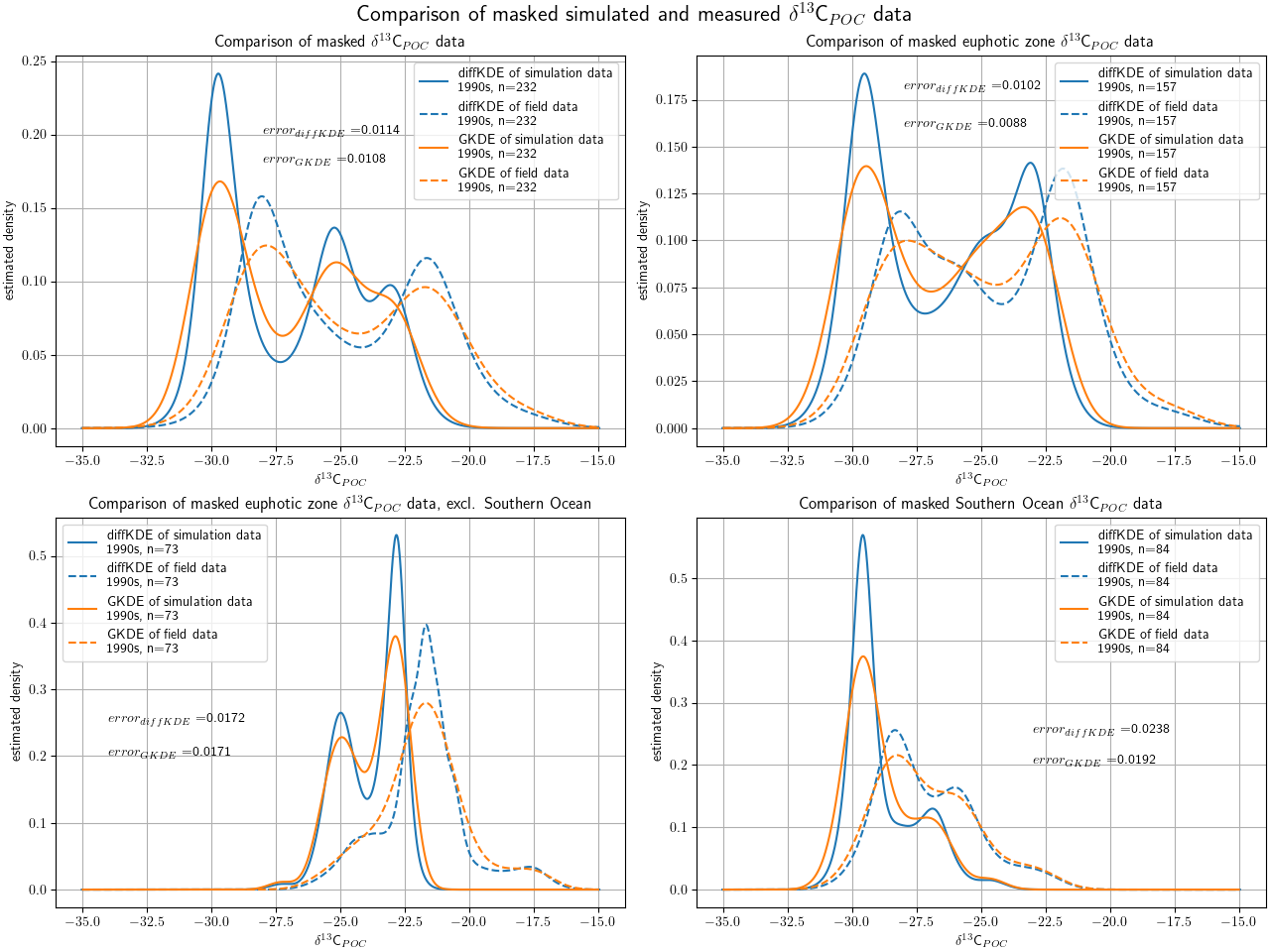}
\caption{Scenario (1) of simulation and field data comparison by comparing KDEs of equally sized data: All four panels show in blue the diffKDE and in orange the Gaussian KDE. The continuous lines are the KDEs calculated from the simulation data. The dashed lines are the KDEs calculated from the field data. All data is averaged over the 1990s and only taken from grid cells, where both data kinds are available. The upper left panel shows the KDEs calculated from all simulation and field data. The upper right panel shows the KDEs calculated from the euphotic zone, i.e. the uppermost 130 m of the oceans. The lower left panel shows the KDEs calculated from the euphotic zones excluding the Southern Ocean, i.e. only data north from $45^{\circ}$ S. The lower right panel shows the KDEs calculated from the euphotic zone of the Southern Ocean, i.e. only data south from $45^{\circ}$ S. The annotated error in each panel is calculated as the Wasserstein distance between simulation and field data for each KDE type. This figure is an adapted version from ~\cite{PelzDissertation}.}\label{fig4}
\end{figure}

\begin{figure}[h]%
\centering
\includegraphics[width=0.9\textwidth]{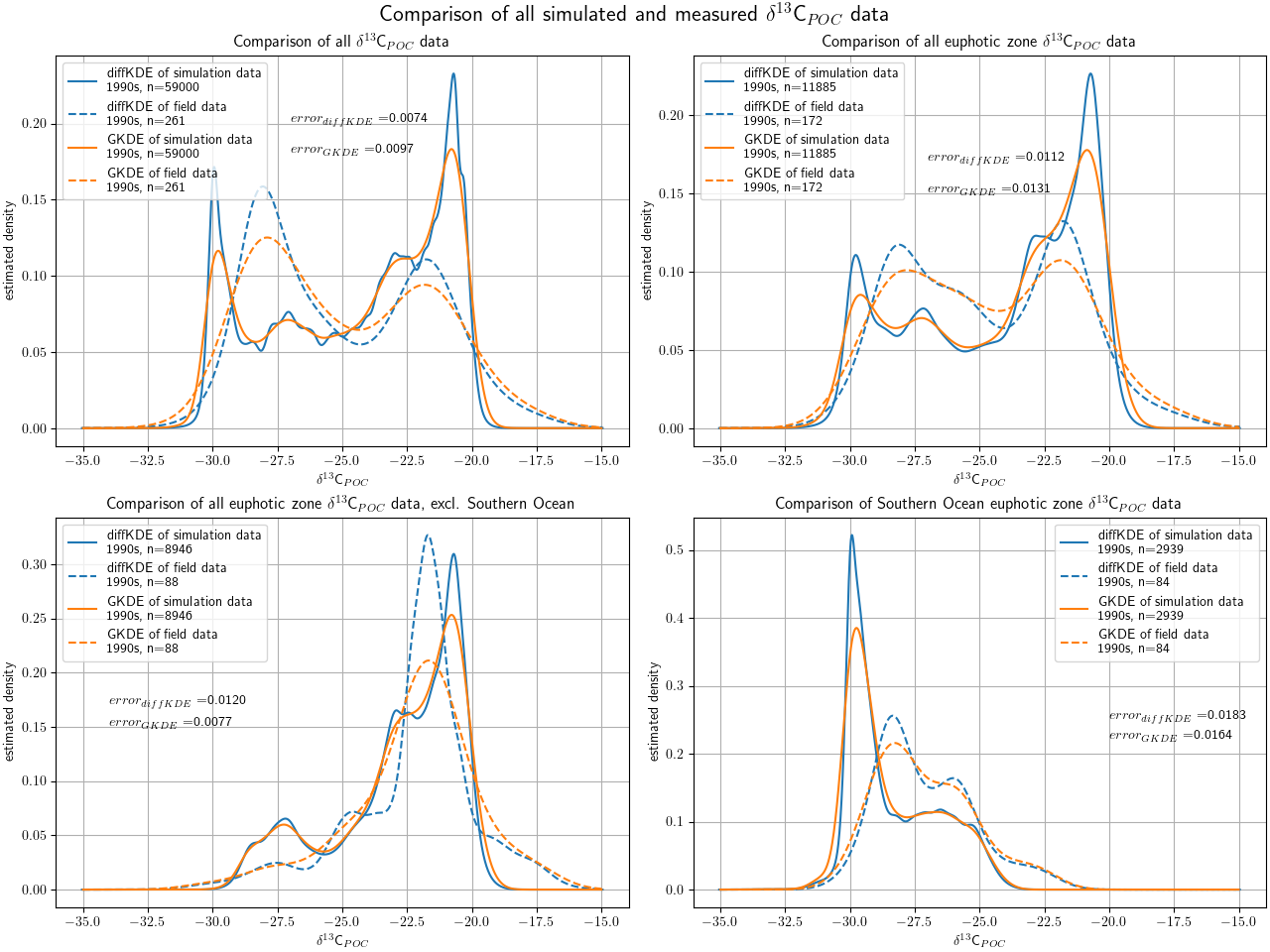}
\caption{Scenario (2) of simulation and field data comparison by comparing KDEs of all available data: All four panels show in blue the diffKDE and in orange the Gaussian KDE. The continuous lines are the KDEs calculated from the simulation data. The dashed lines are the KDEs calculated from the field data. All data is averaged over the 1990s. The upper left panel shows the KDEs calculated from all simulation and field data. The upper right panel shows the KDEs calculated from the euphotic zone, i.e. the uppermost 130 m of the oceans. The lower left panel shows the KDEs calculated from the euphotic zones excluding the Southern Ocean, i.e. only data north from $45^{\circ}$ S. The lower right panel shows the KDEs calculated from the euphotic zone of the Southern Ocean, i.e. only data south from $45^{\circ}$ S. The annotated error in each panel is calculated as the Wasserstein distance between simulation and field data for each KDE type. This figure is an adapted version from ~\cite{PelzDissertation}.}\label{fig5}
\end{figure}

We show the results in Fig. \ref{fig4} and Fig. ~\ref{fig5}. In both figures, all four panels show the comparison of simulation and field data by their estimated densities is shown. The continuous lines present the densities estimated from the simulation results and the dashed lines those estimated from the field data. The blue graphs are estimates by the diffKDE, the orange by the Gaussian KDE. The number of incorporated data points for all is depicted as well as the error calculated by the Wasserstein distance for each KDE approach.

First, we show scenario (1), the assessment based on equally sized data sets. This approach is often employed to compare only data, which has a directly corresponding equivalent in the respective other data set. A typical approach is to incorporate only grid cells into the analyses, where both data kinds are available ~\cite{Somes2021}. By this, one will obtain a specific insight into how well a model simulates tracers at the exactly corresponding location. We realised this approach by using field data interpolated into the same grid as the simulation results ~\cite{pelz2021data} and applying a mask to the data before analyses that only allows the grid cells with both data kinds available.

The overall comparison incorporates 232 data points and shows general too low values in the simulation. Field and simulation data show two main modes each. The lower in the field data is located at around $\delta ^{13}C_{\text{POC}} = -28$ \textperthousand\, and simulated at around $\delta ^{13}C_{\text{POC}} = -29.5$ \textperthousand . The larger mode is located in the field data at around $\delta ^{13}C_{\text{POC}} = -22$ \textperthousand\, and in the simulation results at around $\delta ^{13}C_{\text{POC}} = -25$ \textperthousand . Here, the simulation results reveal an additional less pronounced mode at around $\delta ^{13}C_{\text{POC}} = -23$ \textperthousand\, especially under the diffKDE analysis. Also the pronounciation of the modes is stronger under the simulation data. The lower modes of the field data reaches up to a density of around $0.16$ under the diffKDE and around $0.12$ under the Gaussian KDE. In comparison the corresponding modes reach up to a density of around $0.24$ under the diffKDE and around $0.17$ under the Gaussian. For the higher mode the densities of field data reach up to around $0.12$ under the diffKDE and $0.95$ under the Gaussian, those of simulation data up to $0.14$ and $0.12$ for diffKDE and Gaussian KDE, respectively. The errors between simulation and field data are with $0.0114$ for the diffKDE and $0.0108$ for the Gaussian KDE quite comparable and only slightly smaller under the Gaussian KDE analysis.

The comparison of the euphotic zone data incorporates 157 data points and reveals a similar pattern as the whole ocean comparison. But the second mode is better fit this time. The location of the higher mode is still simulated too low in comparison with the field data. But with around $\delta ^{13}C_{\text{POC}} = -23$ \textperthousand\, the density estimated from the simulation data is a far better fit to the field data in this case. Furthermore, the pronounciation of the higher mode is in good agreement between simulation and field data, now. Also in this case there is an expected third mode under the diffKDE in the simulation data at around $\delta ^{13}C_{\text{POC}} = -25$ \textperthousand\, and in the field data at around $\delta ^{13}C_{\text{POC}} = -26$ \textperthousand\, . The Gaussian KDE does not reveal these patterns. The errors have also reduced and are still comparable, with yet smaller values originating from the Gaussian KDE.

The euphotic zone excluding the Southern Ocean data are 73 data points for each data kind and reveals the simulation of a far too low and more pronounced second lower mode. The main mode is in the field data at around $\delta ^{13}C_{\text{POC}} = -22$ \textperthousand\, and in the simulation data at around $\delta ^{13}C_{\text{POC}} = -23$. A second lower mode is detectable in the field data only under the diffKDE at around $\delta ^{13}C_{\text{POC}} = -24$. The simulation data exhibits a second lower mode far more clearly under both KDEs at around $\delta ^{13}C_{\text{POC}} = -25$. A third mode at around $\delta ^{13}C_{\text{POC}} = -17.5$ \textperthousand\, is visible in the field data under the diffKDE and met by no similar feature in the simulation data. The pronounciations of the modes are again stronger under the diffKDE. The errors are equal  up to one thousand.

The euphotic zone Southern Ocean data are 84 data points for each data kind and their analysis shows again too small values and a far too pronounced smallest mode in the simulation data compared to the field data. The main mode in the field data is located at around $\delta ^{13}C_{\text{POC}} = -28$ \textperthousand\, and a corresponding main mode in the field data at around $\delta ^{13}C_{\text{POC}} = -29.5$. A second smaller mode is located in the field data at around $\delta ^{13}C_{\text{POC}} = -26$ \textperthousand\, and in the simulation data at around $\delta ^{13}C_{\text{POC}} = -27$. In this case and generally for the only time in these examples, the mode in the field data is more pronounced than the mode in the simulation data. A final third mode is visible in the field data only under the diffKDE at around $\delta ^{13}C_{\text{POC}} = -23$ \textperthousand\, and in the simulation data at around $\delta ^{13}C_{\text{POC}} = -24.5$. The pronounciation of the modes is again stronger under the diffKDE. The errors show the biggest difference in this case with a smaller value for the Gaussian KDE.

Second, we use the property of the diffKDE to evaluate unequally sized data in scenario (2) and compare all available data. We use the same four test cases as in the first masked analyses. The data are also the same simulation and field data from ~\cite{Somes2021} and ~\cite{pelz2021data}, respectively, but now all averaged 1990s data are incorporated without the prior restriction to a mask. This leads to the comparison of very differently sized data sets, only restricted to the main location characteristics such as euphotic zone or Southern Ocean describing the four test cases. We show the results in Fig. \ref{fig5}.

The overall data comparison incorporates 59000 simulation and 261 field data points and reveals a better fit of the location of the higher mode between the two data kinds. The lower mode is located similar to the masked example in field an simulation data. The pronounciation of the modes is in better fit now. The higher mode is located in the field data as in the masked analysis. In the simulation data the mode is far higher located now at around $\delta ^{13}C_{\text{POC}} = -21$ \textperthousand\, and by this far better fitting the field data. A third mode in the simulation data is located at around $\delta ^{13}C_{\text{POC}} = -27$ \textperthousand\, in the simulation data only and even a fourth at around $\delta ^{13}C_{\text{POC}} = -23$, mainly under the diffKDE. Some additional smaller structures are visible under the diffDKE in the simulation data between the two outermost modes. The error has decreased for both estimators and is now even smallest under the diffKDE with $0.0074$ in comparison to $0.0114$ in the masked analysis.

The euphotic zone comparison incorporates 11885 simulation data points and 172 data points and reveals similar pattern changes in comparison to the masked comparison as the overall data comparison. The pronounciation of the lower mode is in better fit between the field and simulation data, but with nearly the same offset as in the masked example. The higher mode is located far higher for the simulation data this time at around $\delta ^{13}C_{\text{POC}} = -21$, which leads to a similar offset as in the masked analysis, but now in the opposite direction. The pronounciation of this mode is now too strong in the simulation data in comparison to the field data. There are two additional modes visible in the simulation data at around $\delta ^{13}C_{\text{POC}} = -27$ \textperthousand\, and $\delta ^{13}C_{\text{POC}} = -23$, especially under the diffKDE. A third mode in the field data is expectable at around $\delta ^{13}C_{\text{POC}} = -26$ \textperthousand\, under the diffKDE. The error has slightly increased in this example and is now also smaller for the diffKDE.

The unmasked euphotic zone excluding the Southern Ocean data are 8946 simulation data points and 88 field data points and show a far better fit between both data kinds. The main mode is located at around $\delta ^{13}C_{\text{POC}} = -22$ \textperthousand\, in the field data and at around $\delta ^{13}C_{\text{POC}} = -21$ \textperthousand\, in the simulation data. The pronounciation fits well under the diffKDE and is stronger in the simulation data under the Gaussian KDE. A second and third mode are visible in the field data only under the diffKDE at around $\delta ^{13}C_{\text{POC}} = -25$ \textperthousand\, and $\delta ^{13}C_{\text{POC}} = -27.5$, respectively. The lowest is fit well by a mode in the simulation data, that is more pronounced, but similarly located. A final third mode in the simulation data is again visible only under the diffKDE at around $\delta ^{13}C_{\text{POC}} = -23$. The error has again decreased from the masked analyses to this unmasked analyses for both KDEs and is smaller this time for the Gaussian KDE.

The euphotic zone data restricted to the Southern Ocean incorporates 2939 simulation data points and 84 field data points and shows similar, but better fitting fits between the KDEs as in the masked analysis. The main mode in simulation and field data are similarly located as in the masked example, but this time with a slightly better fitting pronounciation. The location of the second mode is well fitting between simulation and field data in this unmasked analysis, while the pronounciation is similarly different as in the masked analysis. All of this leads to a decreased error, especially for the diffKDE.

\section{Discussion}\label{sec12}

Even highly increased availability of field data leaves these sparse in comparison to simulation results. The here incorporated $\delta ^{13}$C$_{\text{POC}}$ are a highly increased data set version of the at that time latest global version ~\cite{Goericke1994} with nearly 10 times as many globally available data points. Nevertheless, in comparison to simulation results these data are still sparse. This leads to the need to either reduce the data to comparable amounts or to use a comparison measure independent of the available data amount.

Involving all available data improves insights into the model's performance. A decrease of data is inevitably correlated to a loss of information. This is why we chose to employ a comparison measure that allows to take into account all available data. 

Kernel density estimators approximate PDFs from nearly any amount of handed in data. We showed the performance of a classical Gaussian KDE in comparison to the new diffKDE. The diffKDE resolved always more structure of the underlying data and revealed modes that were not detectable under the Gaussian

We conducted four test analyses in different global ocean subsets and see on all four a decrease of the error or at least better qualitative fit for the incorporation of all available data. This mainly comes from a better fit of the location of (one of) the main mode(s). This is generally too low in the density estimated from the masked simulation data. Furthermore, the pronunciation of the modes is often better fitting between unmasked simulation and field data than between the masked data. 

The best improvement of fit is observable in the euphotic zone excluding the Southern Ocean. Location and pronunciation of the main mode are well fitting in the unmasked data comparison, while its pronunciation is overestimated in the masked simulation data and location underestimated in the masked simulation data. The same is true for the second smaller mode, which is even far lower located in the unmasked analysis than in the masked analysis.

Comparison between Gaussian KDE and diffKDE always shows more details revealed under the diffKDE than the Gaussian KDE, leading to a smaller error calculated from the Gaussian KDE. Again, this is most prominent in the Southern Ocean example, where a mode in the field data is detectable under the diffKDE at around $\delta ^{13}C_{\text{POC}} = -27.5$ \textperthousand\, matching a similar one in the simulation data, that is not at all visible under the Gaussian KDE. Furthermore, two additional modes are visible under the diffKDE in the field data at around $\delta ^{13}C_{\text{POC}} = -25$ \textperthousand\, and the simulation data at around $\delta ^{13}C_{\text{POC}} = -23$ \textperthousand\, that are not visible under the Gaussian KDE.

\section{Conclusion}\label{sec13}

The KDE based comparison offers a possibility to compare differently sized data sets as commonly required in model assessment. Such assessment regularly compares simulation results with corresponding field data to validate the model results. This comparison generally requires the comparison of differently sized data, since field data is generally sparse in comparison to globally available simulation data. A common measure is to reduce both data sets to a comparable amount by only incorporating data points from grid cells, where both data kinds are available. This approach naturally leads to a loss of information, which can be avoided by the KDE based approach. The KDE estimates the data's PDF nearly independent of the amount of available data points. The PDFs are then comparable by a proper divergence function such as the Wasserstein distance.

The model resolves general patterns of observational data well. This is well visible under all approaches to model assessment. Number of modes and general shape of the KDEs are mostly comparable.

The traditional approach with the masked data shows too low values from the simulation results in comparison with the field data. This improves a lot when taking into account all available data into the analyses. In this latter approach the error between simulation and field data generally decreases while the fit of location and pronunciation of modes also improves.

The overestimated values in the all data comparison analyses mostly occur in the whole ocean comparison. In the all depths approach as well as in the euphotic zone comparison the higher mode is simulated too high in the model as well its pronunciation too strong in the simulation results.

Mostly underestimated in the all data comparison are the values in the Southern Ocean. The main mode in the Southern Ocean data is too low and too strong pronounced in the simulation data in comparison with the field data. In the whole ocean general and euphotic zone comparison the location of the lower mode is too low in the simulation data, but the pronunciations are comparable.

It is not yet clear what causes the model-data discrepancies. In several cases, simulation and field data seem to fit, but not exactly, as for example in the lower mode in the whole but Southern Ocean data comparison, where the mode is visible in both data kinds but far less expressed in the field data. For future investigations more detailed resolution of the ocean areas can be used to investigate, whether these discrepancies originate from sparse field data in these regions or from actual mismatches simulated by the model.

Overall the comparison by involving all available data provided a more accurate insight into the fit between simulation and field data. This is especially visible for the all but Southern Ocean euphotic zone comparison, where number and location of modes fit significantly better. For future model assessments and calibration this approach can be used to get such accurate insight into model's performance for a variety of applications and very differently sized field data sets.

\section{References}
\bibliography{sn-bibliography}
\bibliographystyle{plain}

\end{document}